\documentclass[letterpaper]{article} 
\usepackage[]{aaai23}  
\usepackage{times}  
\usepackage{helvet}  
\usepackage{courier}  
\usepackage[hyphens]{url}  
\usepackage[hidelinks]{hyperref}
\hypersetup{
    colorlinks,
    linkcolor={red!50!black},
    citecolor={blue!45!black},
    urlcolor={blue!85!black}
}
\usepackage{graphicx} 
\usepackage{natbib}  
\usepackage{caption} 
\usepackage{algorithm}
\usepackage{algorithmic}

\usepackage{newfloat}
\usepackage{listings}
\DeclareCaptionStyle{ruled}{labelfont=normalfont,labelsep=colon,strut=off} 
\floatstyle{ruled}
\newfloat{listing}{tb}{lst}{}
\floatname{listing}{Listing}
\usepackage{subfigure} 
\usepackage{amsmath}
\usepackage[capitalize,nameinlink,noabbrev]{cleveref}
\usepackage{amsfonts}
\usepackage{mathtools}
\usepackage{multirow}
\usepackage{stmaryrd}
\usepackage{ctable}
\usepackage{tcolorbox}
\usepackage{makecell}
\usepackage{bm}

\title{Scaling Law for Recommendation Models: \protect\\Towards General-purpose User Representations}
\author{
    Kyuyong Shin\equalcontrib\textsuperscript{\rm 1}\textsuperscript{\rm 2}~~~
    Hanock Kwak\equalcontrib\textsuperscript{\rm 1}~~~ 
    Su Young Kim\textsuperscript{\rm 1}~~~ 
    Max Nihl\'en Ramstr\"om\textsuperscript{\rm 1}~~~  \\
    Jisu Jeong\textsuperscript{\rm 1}\textsuperscript{\rm 2}~~~ 
    Jung-Woo Ha\textsuperscript{\rm 1}\textsuperscript{\rm 2}~~~ 
    Kyung-Min Kim\textsuperscript{\rm 1}\textsuperscript{\rm 2}
}
\affiliations{
    \vspace{1mm}
    \textsuperscript{\rm 1}NAVER CLOVA~~
    \textsuperscript{\rm 2}NAVER AI Lab  \\
    \vspace{1mm}
    \small\{ky.shin, hanock.kwak2\}@navercorp.com
}

\begin{document}
\maketitle
\begin{abstract}
Recent advancement of large-scale pretrained models such as BERT, GPT-3, CLIP, and Gopher, has shown astonishing achievements across various task domains. Unlike vision recognition and language models, studies on general-purpose user representation at scale still remain underexplored.
Here we explore the possibility of general-purpose user representation learning by training a universal user encoder at large scales. We demonstrate that the scaling law is present in user representation learning areas, where the training error scales as a power-law with the amount of computation.
Our \textbf{C}ontrastive \textbf{L}earning \textbf{U}ser \textbf{E}ncoder (CLUE), optimizes task-agnostic objectives, and the resulting user embeddings stretch our expectation of what is possible to do in various downstream tasks. 
CLUE also shows great transferability to other domains and companies, as performances on an online experiment shows significant improvements in Click-Through-Rate (CTR). Furthermore, we also investigate how the model performance is influenced by the scale factors, such as training data size, model capacity, sequence length, and batch size. Finally, we discuss the broader impacts of CLUE in general.
\end{abstract}
\section*{Introduction}
Recent work has demonstrated that models pretrained on enormous data at scale can perform remarkable downstream transfers in a flexible and task-agnostic manner for vision recognition~\cite{zhai2021scaling,dai2021coatnet,goyal2021self}, language models~\cite{devlin2018bert,NEURIPS2020_1457c0d6,kim2021changes,rae2021scaling}, speech recognition~\cite{baevski2020wav2vec}, and multimodal learning~\cite{radford2021learning,akbari2021vatt,wang2021simvlm}. 
These large-scale models are called \textit{foundation models}, which have brought seismic changes to both academia and industry by providing general-purpose utilities and promising results~\cite{bommasani2021opportunities}. 
However, in contrast to the other domains mentioned above, studies on the scaling and generalization ability of general-purpose pretrained user representation learning still remain underexplored for various downstream recommendation tasks.
\begin{figure*}[t]
\begin{center}
\centerline{\includegraphics[width=0.80\textwidth]{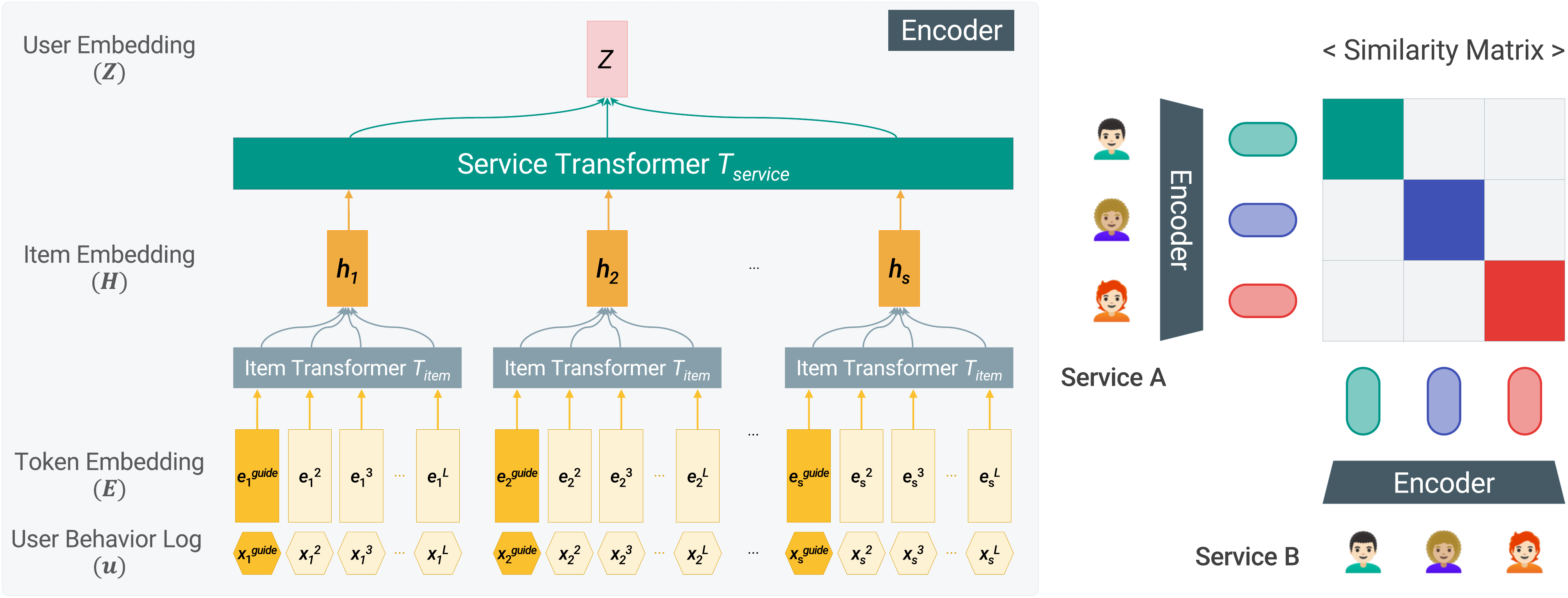}}
\caption{Overall flow of CLUE. The encoder encodes user behavior logs, a sequence of items described in natural language text ($u$). This description becomes token embeddings ($E$) and then passed through the Item Transformer followed by mean pooling ($H$). The sequence of item embedding vectors are then passed to the Service Transformer. The user embedding ($Z$) is the mean of the last hidden vectors of the Service Transformer. Finally, the user embeddings are contrastively pretrained.}
\label{clue_summary}
\end{center}
\end{figure*} 

Here we highlight five key questions for exploring the possibilities of general-purpose user representations.
1) Can general-purpose user representations learned from multiple source data provide a promising transfer learning capability?
2) Are pretraining and downstream task performances positively correlated?
3) How various tasks can the pretrained user representations address?  
4) Does scaling up the pretraining model improve the generalization performance? 
5) If so, which factors, such as, training data and model size, behavior sequence length, and batch size, should be scaled up?

To answer these questions, we introduce \textbf{C}ontrastive \textbf{L}earning \textbf{U}ser \textbf{E}ncoder, i.e, CLUE. CLUE demonstrates the effectiveness of the pretrained general user representations, learned from 50 billion behavior tokens of 11 million users from a search engine and e-commerce platform that share a common user pool, by adapting them to unseen downstream tasks.
CLUE uses contrastive learning by constructing user representations for each service task, then treating pairs of the same user representations as positive samples while treating representations of different users as negative samples.
We comprehensively evaluate the pretrained user representation of CLUE with multiple downstream tasks from industrial and benchmark datasets, including an online CTR evaluation. More specifically, we compare the performance of a simple multi-layer perceptron (MLP) employing our task-agnostic pretrained CLUE features with a task-specific model trained for each downstream task. Furthermore, we investigate the empirical scaling laws of training data size, model size, sequence length and batch size with extensive experiments, and analyze power-law scaling for training performance as a function of computing resources.

The summary of our key findings through CLUE is as follows:

\noindent\textbf{Empirical scaling law.}
The pretraining test error scales as a power-law with the total training cost (PF-days) unless bottlenecked by other factors, which is also observed in other domains~\cite{NEURIPS2020_1457c0d6}. However, we observe that the performance of the models does not depend on their size alone in user representation tasks (\cref{fig:comp_a}), other factors such as batch size and sequence length also have considerable influences on the performance. 

\noindent\textbf{Transferability improves as the pretraining error decrease.}
CLUE performs transfer learning well on heterogeneous datasets even when the domains are different.
The resulting test losses on the downstream tasks also show strong correlation to the pretraining test loss. These results indicate that generalization on various data distributions are strongly dependent on the pretraining error (\cref{transfer}-Right).

\noindent\textbf{Transforming tabular data to natural language text provides common semantic representation.} We transform all data into natural language texts by extracting textual information from tabular data (e.g., product descriptions from product data table). 
This policy alleviates the discrepancies in data format within different services; the data format of the same product varies depending on the platform, but the product name is still the same.
We show that CLUE trained on data from a particular company can provide meaningful user representations for another company. 

\noindent\textbf{Advantages of training from multiple service logs.}
CLUE learns a multi-modal user embedding space from two services, and shows promising results on diverse downstream tasks. Furthermore, the results (\cref{tab:online}) demonstrate that pretraining on multiple services effectively addresses the cold-start problem by learning a richer user representation.

\section*{CLUE}
Performing general representation learning on training data from different services and producing a unified view is a crucial challenge in building foundation models for user modeling.
In this paper, we look into a multi-modal contrastive learning framework for enhancing the representation quality of the user embeddings by considering each service as a modality.
A notable example of a multi-modal contrastive learning framework is CLIP~\cite{radford2021learning}, which encodes pairs of images and texts separately into vectors and maximize their similarity scores. CLIP outperforms the best publicly available models in a wide range of computer vision tasks. 

A remaining issue is aligning user semantics across multiple services. While CLIP leverages image-text paired data for semantic alignment, it is non-trivial to align the items of two heterogeneous services. To address this issue, we use the text description of each item rather than its ID as inputs. Natural language input for items enables our user representations to naturally and flexibly transfer to other services and even other company platforms where items can be described with text, thus enhancing the generality of our method. As illustrated in \cref{clue_summary}, such a flexible embedding space allows us to jointly train user logs over multiple services, such as search engine and e-commerce, by maximizing the agreement between the same user with a single encoder.

Our encoder has a hierarchical structure consisting of an Item Transformer, $T_{\text{item}}$, and a Service Transformer, $T_{\text{service}}$. The Item Transformer encodes the features of each item in the user's behavior log, namely $T_{\text{item}}: \mathbb{R}^{L\times D_{\text{in}}}\rightarrow\mathbb{R}^{L\times D_{\text{out}}}$. The Service Transformer encodes a user embedding for a specific service from the sequence of item embeddings, i.e. $T_{\text{service}}: \mathbb{R}^{S\times D_{\text{out}}}\rightarrow\mathbb{R}^{S\times D_{\text{out}}}$. Here $D_{\text{in}}$, $D_{\text{out}}$, $L$ and $S$ denote the token embedding dimension, the output dimension, the maximum token length of an item description, and the number of items in an item sequence, respectively.
For each service, we define a user behavior log as $u = [\bm{x}_1, \bm{x}_2, \ldots, \bm{x}_S]$, a sequence of items that a user acts upon or selects in the service.
Each item $\bm{x}_i \in \mathcal{V}^{L}$ is a vector of token indices where the first couple tokens represent the type of service and the following tokens are the tokenized description of the item (e.g., product description, news title, or search query).
We fill the remaining spaces with zeros (see more details in \cref{sec:app-dataset}).
Each token is embedded  to a vector via an embedding layer $g: \mathcal{V}\rightarrow\mathbb{R}^{D_{\text{in}}}$, so each item $\bm{x}_i$ is embedded to a matrix $E_i \in\mathbb{R}^{L\times D_{\text{in}}}$. The token embedding $E_i$ is then propagated through the model in the following manner:
\begin{align} 
\bm{h}_{i} = \texttt{MEAN}(T_{\text{item}}(E_{i})),\\ 
H = [\bm{h}_1 | \bm{h}_2 | \cdots | \bm{h}_S],\\ 
\bm{z} = \texttt{MEAN}\left(T_{\text{service}}(H)\right), 
\end{align}
where \texttt{MEAN} is mean pooling of row vectors in the input matrix, and $\bm{z}$ is the final user embedding.
We again denote the final user embedding of user $u$ and service $A$ as $\bm{z}_{u, A}$.
We follow the loss of CLIP~\cite{radford2021learning}.
The loss $l_{u,A,B}$ of each positive pair $(\bm{z}_{u,A}, \bm{z}_{u,B})$ is defined as:
\begin{equation} 
\centering
l_{u, A, B} = -log\frac{\text{exp}(\big<f(\bm{z}_{u,A}), f(\bm{z}_{u,B})\big>\tau)}
{\sum_{v} \text{exp}(\big<f(\bm{z}_{u,A}), f(\bm{z}_{v,B})\big>\tau)},
\end{equation}
where $\tau$ is a temperature parameter and $\big<\cdot, \cdot\big>$ is the cosine similarity.
We use a non-linear projection model $f(\bm{z})=W_{2}\sigma(W_{1}\bm{z})$ to improve the representation quality of the user embeddings~\cite{chen2020simple}, where $\sigma$ is a non-linear activation function and $W$ is a weight matrix. We optimize the symmetric cross-entropy loss $(l_{u, A, B} + l_{u, B, A})/2$. The final user features for the downstream tasks are extracted by concatenating each service user feature, or for the case of the company-level transferability task, extracted by using only task-specific user logs. 

The training details and hyperparameters of best CLUE are described in \cref{sec:app-training}.

\section*{Experiments}
\subsection{Pretraining Dataset}
We construct a sufficiently large-scale dataset with more than 50B behavior tokens collected over 2 years from search engine and e-commerce platform. We exclude the users who act less than once every two months in terms of behavior log frequency. We use Byte-level BPE (BBPE)~\cite{wang2020neural} to tokenize the textual description of each item in the user behavior logs. 
If a user repeated the same behavior (e.g., performed the same search query or purchased the same product multiple times), we keep only one of the entries in the behavior log, to count it as a unique behavior.   
As a result, the training dataset contains 11 million users and 5.3 billion user behavior logs, and 50 billion BBPE tokens collected over 2 years. The 11 million users are sampled for training, but any user can be a candidate for the downstream tasks. The statistics of the dataset for pretraining are provided in \cref{sec:app-dataset}.

\subsection{Downstream Tasks}
\noindent\textbf{Benchmark dataset.} 
We select two categories \textit{“Books”} and \textit{“Clothing Shoes and Jewelry”} from Amazon review dataset~\cite{ni2019justifying}. 
This dataset contains product reviews and scores with product metadata like product titles and categories. We filtered the case where the review score was 4 points or higher, and only the title of the product was used as metadata. 
Each user's review history was listed chronologically and used as a historical log, and up to 64 were used. The most recent three reviews were used as target sets.

\noindent\textbf{Industrial dataset.} 
We build downstream tasks using data from services that are different from the service domains used in the pretraining dataset. For example, downstream tasks are built from services of an e-commerce (PCR), web-based cartoon (FWR), news (NVR), marketing messages (MMR), and an online travel agency (OTAR). Furthermore, we validate the company-level transferability of our model, defined as Inter-Company-Level Transfer (ICLT). To achieve this, we secure log data from a completely different company that has an online marketplace environment (see more details in \cref{sec:app-iclt}). We describe details of the downstream datasets in \cref{sec:app-downstream}.

\noindent\textbf{Experimental Settings.} 
The downstream tasks are composed of recommendation tasks where the models predict the next item to recommend.
The datasets consist of positive and negative pairs $(u, i)$ of users and items.
We have a positive pair $(u, i^{+})$ when a user $u$ interacted with an item $i^{+}$, while a negative pair $(u, i^{-})$ is generated through random sampling.
We compute three standard metrics for evaluation: top-$k$ Hit Ratio (HR@$k$), top-$k$ Normalized Discounted Cumulative Gain (NDCG@$k$), and Mean Reciprocal Rank (MRR). We obtain these metrics by evaluating a pool of items consisting of a ground-truth item mixed with 100 randomly sampled negative items. To test the generalization ability of the models, we make sure there are no shared users between the training, validation, and test sets.

\setlength{\tabcolsep}{2pt}
\ctable[
    caption = {Computational cost comparison of the downstream models measured from the Books task.},
    label = tab:resource,
    pos=t,
      doinside=\small
]{cccccc}{
\tnote[$\dagger$] {Transformer based models including BST, UserBERT, and UniSRec}}{
\toprule
Models & Inputs & Speedup & Params. & Memory \\
\midrule
Transformer$^\dagger$ & Task-specific logs & 1 & 15M & 1G \\
LightGCN & Task-specific logs & 10$\times$ & 0.5M & 4G \\
CLUE & Pretrained user repr. & 43$\times$ & 0.5M & 0.5G \\
\bottomrule
}

\setlength{\tabcolsep}{12.0pt}
\ctable[
    caption = {Results on the benchmark downstream tasks. 
    Pretraining and then transferring models (i.e., UniSRec, UserBERT, and CLUE) are pretrained using the history logs of “Books”
and “Clothing Shoes and Jewelry”. The user feature from CLUE is extracted as same as the ICLT task. The best results are underlined.},
    label = tab:public,
    pos=t,
    star,
    doinside = \scriptsize
]{ccccccccc}{
}{
\toprule 
Downstream tasks & Metrics & DeepFM & BST & LightGCN & YTMoE & UserBERT & UniSRec & CLUE (15M)  \\
\midrule[0.52pt]
\midrule[0.52pt]
\multirow{3}{*}{\makecell{Books}} 
& HR@1    & 0.0580 & 0.0676 & 0.0889 & 0.0947 & 0.0850 & 0.0618 & \underline{0.1087}  \\
& NDCG@10 & 0.1857 & 0.1916 & 0.1875 & 0.1925 & 0.1826 & 0.1895 & \underline{0.2104}  \\
& MRR     & 0.1552 & 0.1606 & 0.1690 & 0.1742 & 0.1644 & 0.1601 & \underline{0.1854}  \\
\midrule
\multirow{3}{*}{\makecell{Clothing}} 
& HR@1    & 0.0812 & 0.1169 & 0.0924 & 0.0846 & 0.1208 & 0.1091 & \underline{0.1564}  \\
& NDCG@10 & 0.2082 & 0.2294 & 0.2245 & 0.2048 & 0.2544 & 0.2228 & \underline{0.2857}  \\
& MRR     & 0.1777 & 0.2046 & 0.1929 & 0.1783 & 0.2228 & 0.1971 & \underline{0.2481}  \\
\bottomrule
}
\setlength{\tabcolsep}{6.5pt}
\ctable[
    caption = {
    Results on the industrial downstream tasks. 
    The best results among the Task-agnostic and all models are denoted in bold and underlined fonts, respectively.
},
    label = tab:downstream-tasks,
    pos=t,
    star,
    doinside = \scriptsize
]{ccccccccccc}{
\tnote[$\dagger$] {Hybrid for \textbf{task-specific (BST or LightGCN)} models enhanced with CLUE features.}
}{
\toprule 
\multirow{2}{*}{Downstream tasks} 
& \multirow{2}{*}{Metrics} 
& \multicolumn{5}{c}{Task-specific} 
& \multicolumn{3}{c}{Task-agnostic}  
& \multirow{2}{*}{Hybrid} 
\\
\cmidrule(l){3-7}  
\cmidrule(l){8-10}  
& & DeepFM & BST & LightGCN & UserBERT & UniSRec & ShopperBERT & SimCLR & CLUE  \\
\midrule[0.52pt]
\midrule[0.52pt]
\multirow{3}{*}{\makecell{PCR}} 
& HR@1    & 0.4297 & 0.4832 & 0.4792 & 0.5134 & 0.5024 & 0.4991 & 0.5114 & \textbf{0.5414} & \underline{0.5485} \\
& NDCG@10 & 0.6696 & 0.7017 & 0.6988 & 0.7140 & 0.7126 & 0.7174 & 0.7231 & \textbf{0.7418} & \underline{0.7466} \\
& MRR     & 0.5979 & 0.6380 & 0.6344 & 0.6642 & 0.6478 & 0.6552 & 0.6626 & \textbf{0.6857} & \underline{0.6912} \\
\midrule
\multirow{3}{*}{\makecell{MMR}} 
& HR@1    & 0.1609 & 0.2498 & 0.2485 & 0.2874 & 0.2648 & 0.2744 & 0.3032 & \textbf{\underline{0.3146}} & 0.2987 \\
& NDCG@10 & 0.4430 & 0.5354 & 0.5325 & 0.5627 & 0.5504 & 0.5637 & 0.5873 & \textbf{\underline{0.6030}} & 0.5856 \\
& MRR     & 0.3048 & 0.4019 & 0.3995 & 0.4514 & 0.4326 & 0.4298 & 0.4578 & \textbf{\underline{0.4713}} & 0.4532 \\
\midrule
\multirow{3}{*}{\makecell{NVR}} 
& HR@1    & 0.5492 & 0.5750 & 0.4412  & 0.5790 & 0.5815 & 0.5492 & 0.5720 & \textbf{0.5927} & \underline{0.5950} \\
& NDCG@10 & 0.6992 & 0.7161 & 0.5849  & 0.7127 & 0.7108 & 0.7025 & 0.5589 & \textbf{0.7253} & \underline{0.7287} \\
& MRR     & 0.6615 & 0.6806 & 0.5498  & 0.6866 & 0.6903 & 0.6665 & 0.6789 & \textbf{0.6924} & \underline{0.6954} \\
\midrule
\multirow{3}{*}{\makecell{OTAR}} 
& HR@1    & 0.1654 & 0.1616 & 0.1794 & 0.1628 & 0.1720 & 0.1739 & 0.1754 & \textbf{0.1977} & \underline{0.2003} \\
& NDCG@10 & 0.4097 & 0.4092 & 0.4235 & 0.4045 & 0.4161 & 0.4192 & 0.4216 & \textbf{0.4445} & \underline{0.4471} \\
& MRR     & 0.3285 & 0.3292 & 0.3439 & 0.3334 & 0.3429 & 0.3393 & 0.3415 & \textbf{0.3653} & \underline{0.3682} \\
\midrule
\multirow{3}{*}{\makecell{FWR}} 
& HR@1    &  &                 &  &   &   & 0.1281 & 0.1282 & \textbf{\underline{0.1538}} &  \\
& NDCG@10 & \multicolumn{5}{c}{No History Logs}  & 0.7015 & 0.7003 & \textbf{\underline{0.7220}} & No History Logs \\
& MRR     &  &                 &  &   &   & 0.2464 & 0.2468 & \textbf{\underline{0.2804}} &    \\
\midrule
\multirow{3}{*}{\makecell{ICLT}} 
& HR@1    & 0.4109 & 0.4489 & 0.4781 & 0.4891 & 0.4915 &     & 0.4543 & \textbf{0.5081} & \underline{0.5262} \\
& NDCG@10 & 0.6302 & 0.6620 & 0.6832 & 0.7017 & 0.6982 & N/A & 0.6628 & \textbf{0.7036} & \underline{0.7112} \\
& MRR     & 0.5631 & 0.5964 & 0.6215 & 0.6334 & 0.6312 &     & 0.5987 & \textbf{0.6440} & \underline{0.6560} \\
\bottomrule
}

\subsection{Downstream Models}
We have two types of downstream models, \textbf{task-agnostic} and \textbf{task-specific} models. \textbf{Task-agnostic} models use the pretrained user representation as the user feature and project it with a simple MLP (input-512-256-128-64-output, ReLU) for each downstream task. CLUE falls into this category of models, and we compare its performance to two other task-agnostic models, ShopperBERT~\cite{shin2021one4all} and SimCLR-based~\cite{chen2020simple} learning, to validate the effectiveness of our proposed learning strategy. \textbf{Task-specific} models---all the models except ShopperBERT, SimCLR, and CLUE---use task-specific historical logs of users as the user features. In addition, we consider a \textbf{Hybrid} model that combines the task-specific user historical logs with the pretrained user features from CLUE, for additional insight. For the item embeddings of the downstream tasks, we utilize the item's text information. We use Sentence-BERT~\cite{reimers2019sentence} to extract item embedding vectors for the \textbf{task-specific} models, while we use the task-agnostic pretrained models (i.e., ShopperBERT, SimCLR and CLUE) for the \textbf{task-agnostic} models. 

For the \textbf{task-agnostic} models, we use two different MLPs for projecting the item features and the user features. The $logits$ are calculated as the dot product between the projected user features and the item features.
Note that using raw task-specific logs requires much more computational costs
(see \cref{tab:resource}). The detailed description of comparison models are outlined in \cref{sec:app-comparison}

\begin{figure*}[ht]
\centering
\subfigure[Model Size vs. Batch Size vs. Sequence Length]{\label{fig:comp_a}\includegraphics[width=0.46\textwidth]{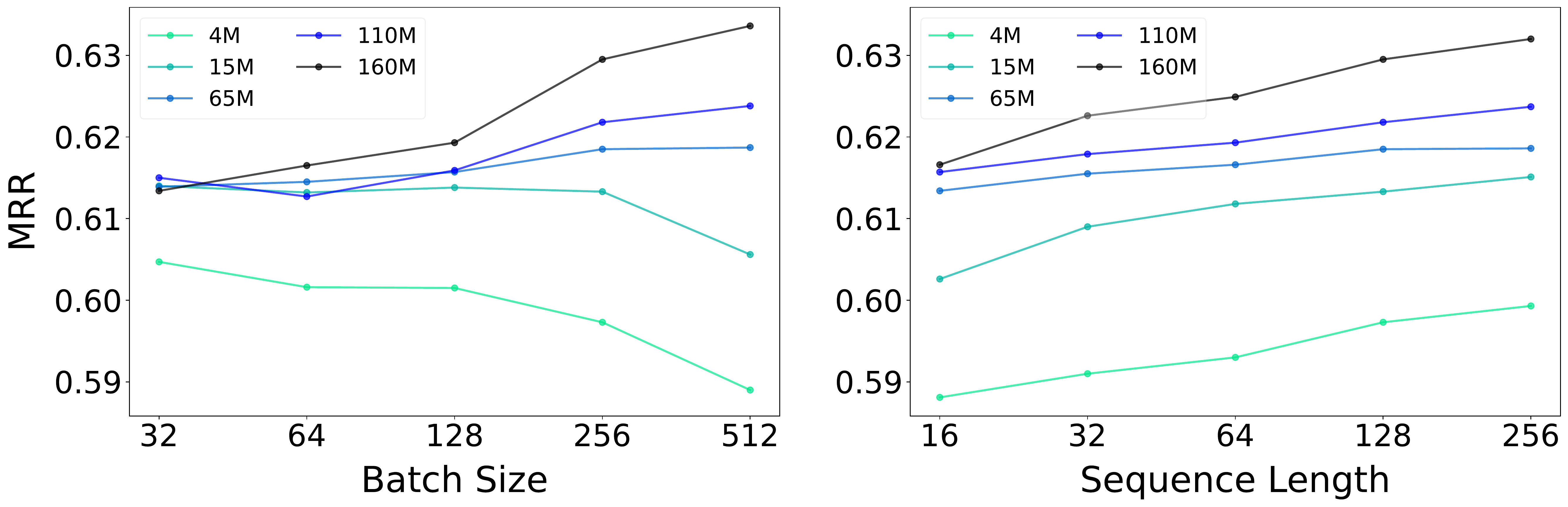}}
\subfigure[ICLT performance according to the amount of computation]{\label{fig:comp_b}\includegraphics[width=0.46\textwidth]{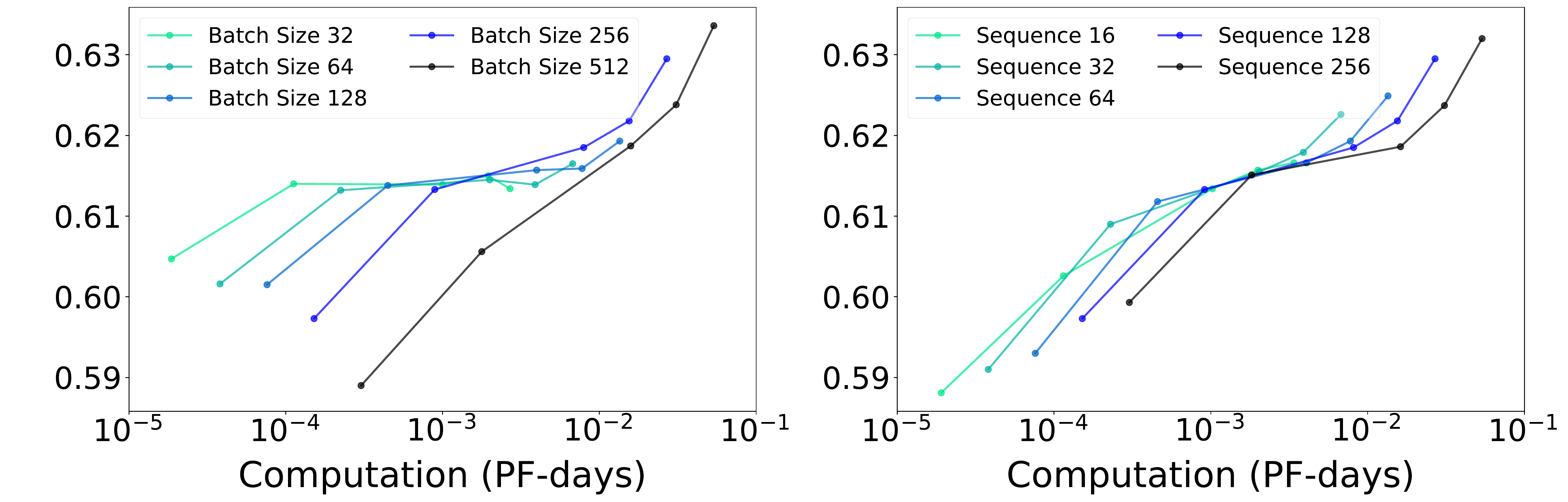}}
\vspace{-0.15in}
\caption{As more computing resources become available, we can choose how to allocate the resources when scaling up the training: larger models, larger batches, and longer sequence lengths. (a) The performance on the ICLT task when either the sequence length (128) or the batch size (256) is fixed. (b) To evaluate the efficiency of the training schemes, we report the performance improvement in terms of the amount of training computation (PF-days). Within the same line, each dot represents a model size ranging from 4M- to 160M-parameters in increasing order.}
\label{batch-dataset-model-comp}
\end{figure*}
\begin{figure}[ht]
\begin{center}
\centerline{\includegraphics[width=0.93\columnwidth]{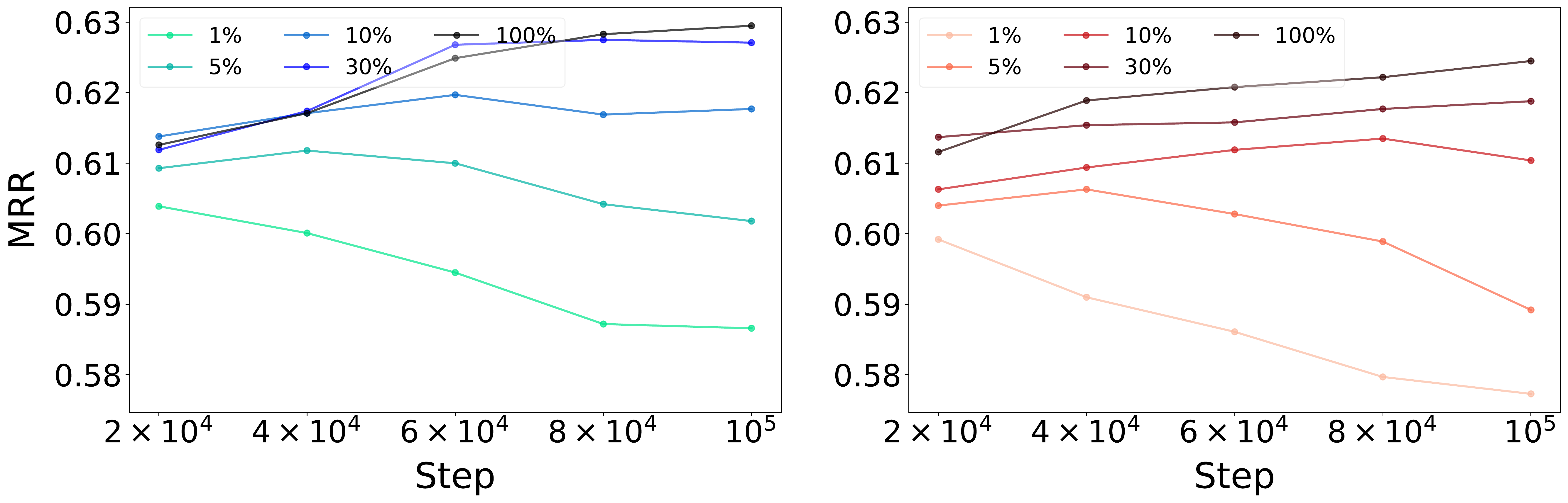}}
\caption{Performance on the ICLT task according to the number of training steps while varying the size of the training dataset from $1\%$ to $100\%$. CLUE is trained with 160M parameters, sequence length (128), and batch size (256). (Left) With batch shuffling during training to help the model learn from the various negative examples. (Right) Without shuffling during training.}
\label{steps}
\end{center}
\end{figure} 
\setlength{\tabcolsep}{3.5pt}
\ctable[
    caption = {
    Performance gain on the online PCR task, compared with a baseline method ($\text{TopPop}^{\dagger}$). 
},
    label = tab:online,
    pos=t,
      doinside=\small
]{ccccc}{
\tnote[$\dagger$] {TopPop recommends the most popular items to any user, regardless of their preferences.}
}{
\toprule \multirow{2}{*}{Method} & \multicolumn{4}{c}{Engagement Metric}  \\
\cmidrule(l){2-5}  & New & Cold & Heavy & Total  \\
    \midrule[0.52pt]
    \midrule[0.52pt]
    GNN~\cite{jeong2020div2vec}     & -0.7$\%$   & +9.0$\%$   & +9.8$\%$   & +2.9$\%$  \\
    CLUE (15M)  & +4.1$\%$   & +11.0$\%$   & +10.5$\%$   & +6.5$\%$  \\
    CLUE (120M) & +4.5$\%$   & +13.4$\%$   & +10.7$\%$   & +7.3$\%$  \\
\bottomrule
}
\section*{Results}
\subsection{User Representations for Downstream Transfer} 
\label{sec:down-result}
We present the adaptability and generality of the pretrained user representations by showing state-of-the-art performance on diverse services, which data has different distributions from our pretraining data. 
In addition, we show that the simple transfer learning using CLUE remarkably outperforms the complex models in two benchmark datasets. 
This empirically demonstrates the generalization ability of the pretrained features by CLUE in wide spectrum of applications. The results are presented in \cref{tab:public}, \ref{tab:downstream-tasks}, and~\ref{tab:online}.

\noindent\textbf{Results on the Benchmark datasets.}
We compare CLUE with six task-specific models, including the models train from scratch---DeepFM~\cite{guo2017deepfm}, BST~\cite{chen2019behavior}, LightGCN~\cite{he2020lightgcn}, and YTMoE~\cite{zhao2019recommending}---and pretraining and then finetuning models, i.e., UniSRec~\cite{hou2022towards} and UserBERT~\cite{wu2021userbert}. For all the tasks and metrics, CLUE outperforms all models by meaningful margins, particularly on MMR (over 6\%). These results show effectiveness of our multi-service contrastive learning with a feature-based transfer framework compared to other seq-to-seq contrastive models with finetuning even with considerable computation efficiency (\cref{tab:resource}).

\noindent\textbf{Results on the Industrial datasets.}
CLUE significantly outperforms the other methods in all tasks and metrics. For the Product Collection Recommendation and Marketing Message Recommendation tasks, CLUE outperforms best baseline models by over $3\%$ in terms of MRR.
CLUE achieves 0.3653 and 0.2804 MRR scores on Online Travel Agency Recommendation and Favorite Webtoon Recommendation, that is a $6\%$ and $14\%$ increase compared to the MRR score of best basline models, respectively. 
In the News View Recommendation tasks, CLUE shows the only marginal performance improvement compared to UniSRec.

For the Inter-Company-Level Transfer task, the results of CLUE on the MRR metric is higher than that of UserBERT by $2\%$. ShopperBERT cannot be evaluated on the ICLT task due to its product ID-based MLM loss, since the target company uses different product IDs from our system.

It is worth noting that the transfer learning with the user representations by CLUE results in consistent performance gains in challenging real-world downstream applications, showcasing the effectiveness of the proposed CLUE-based user representation learning. These results support the efficacy of our CLIP-style multi-service contrastive learning. 

Also, we report the results of a hybrid approach combining user features from CLUE and a task-specific model with historical logs. This hybrid shows the best performances on most tasks and metrics except for MMR. These results support our CLUE can be complementary with task-specific models and improve the conventional methods by providing generalized user features. 

\noindent\textbf{Online Results.}
We conducted an online test for PCR task on our e-commerce platform for five days in November 2021. We split the users into three groups---new, cold, and heavy---based on their engagement frequency for each service. 
The user group `new' corresponds to users with no recorded behavior on the service for the past month. The user group `cold' corresponds to the bottom 10$\%$ in terms of the amount of activity while the group `heavy' represents the top 10$\%$. 
The results are presented in \cref{tab:online}.

Compared with the conventional baseline TopPop, CLUE (120M) significantly increased the Click-Through-Rate (CTR) engagement by $7\%$. As a task-specific model, we employ a graph neural network (GNN)~\cite{jeong2020div2vec} which is a random-walk based graph representation learning method. CLUE (120M) comfortably outperforms GNN by $5\%$ in terms of CTR engagement. 
It is worth mentioning that CLUE (120M) obtains more CTR engagement than CLUE (15M). The result verifies that the universal scaling law still works in online scenarios. 

Furthermore, we investigate how the models perform with different user segments. For user groups `new' and `cold', the user representation pretrained by CLUE significantly improves CTR compared to GNN. In particular, the CTR of GNN is lower than that of TopPop for the `new' user group that lacks behavior logs, while CLUE consistently shows outstanding performances. Interestingly, as the amount of task-specific history logs increases from `new' group  to `heavy' group, the CTR difference between GNN and CLUE decreases. We conjecture that CLUE might provide more generalized user representation for new and cold users, thus contributing to practical applications.

\begin{figure}[t]
\begin{center}
\centerline{\includegraphics[width=0.95\columnwidth]{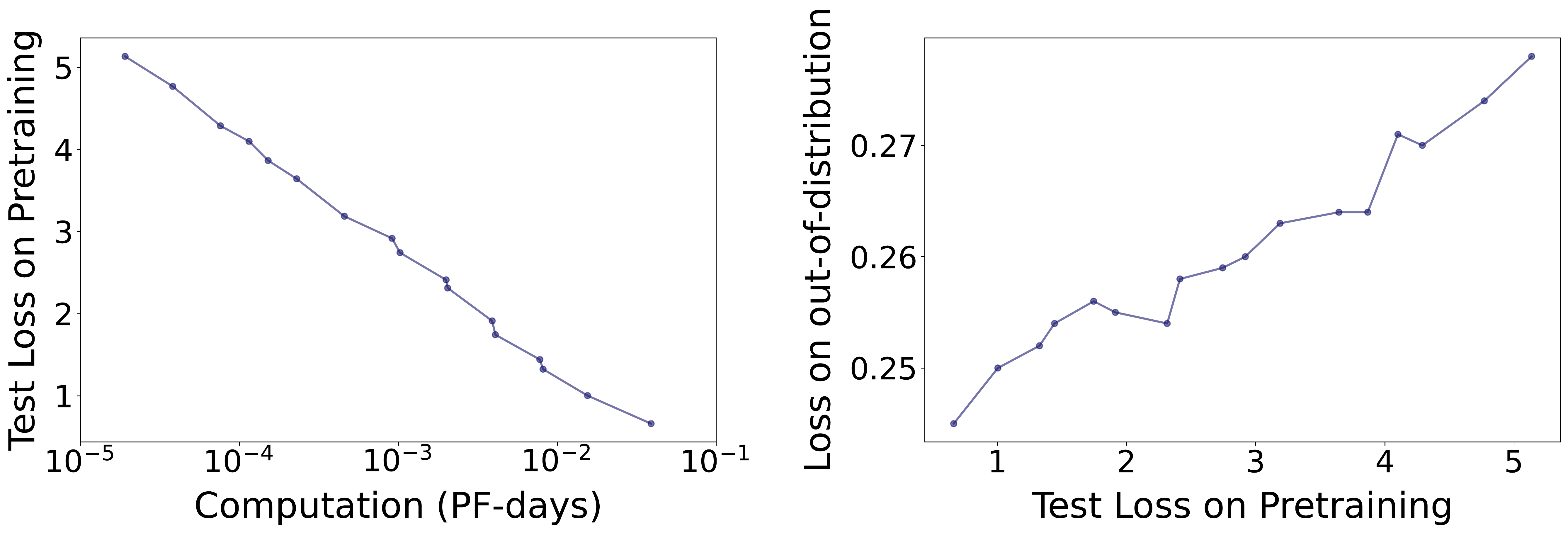}}
\caption{(Left) Log-linear plot between the test loss on the pretraining distribution and the computation (PF-days). This shows that the pretraining learning curve has a power-law scaling as a function of the total training computation (PF-days) when not bottlenecked by other factors. (Right) The generalization on downstream tasks depends on the  test loss of pretraining. We observe a strong trend that a lower loss on the in-distribution data results in a lower loss on the out-of-distribution data.}
\label{transfer}
\end{center}
\end{figure} 
\subsection{Scaling Laws and Generalization} \label{sec:scale}
Recently, several lines of work empirically demonstrate the existence of a scaling law, where the training error scales as a power-law with model capacity, data size, and the amount of computation~\cite{NEURIPS2020_1457c0d6, kaplan2020scaling, zhai2021scaling, bahri2021explaining}. Studies on the scaling law have significantly broadened the field of reasoning, but the findings are mostly restricted to NLP and computer vision. In this subsection, we empirically observe the power-law decrease of the pretraining error in user modeling areas w.r.t. scales. The computation (PF-days) is calculated as 6 $\times$ \verb|#| of parameters $\times$  batch size $\times$ \verb|#| of training steps $\times$ sequence length divided by one PF-day = $8.64\times10^{19}$. We train  all models for 100,000 steps. 

\noindent\textbf{Performance on Scale.}
In contrast to existing work on the scaling law in other domains, we observe that the performance of the model does not depend dominantly on model capacity alone in the user representation tasks (\cref{fig:comp_a}-Left). Previous studies argued that learning high-quality representations from batch-wise contrastive loss requires a sufficient amount of negative samples~\cite{chen2020simple, mitrovic2020representation, gao2021scaling}. Thus, we speculate that the scaling law when learning with a contrastive objective is more complex than that of supervisory signals due to the bottleneck induced by the batch size.

CLUE’s performance grows smoothly as the sequence length of the input data increases, suggesting that user behavior models benefit from observing longer sequences of customer behavior (\cref{fig:comp_a}-Right).
We can see the performance change according to the amount of training computation (\cref{fig:comp_b}). These results give us an insight into how to appropriately allocate computing resources for efficient training schemes. 

We further conduct an experiment on increasing the training dataset size from $1\%$ to $100\%$. \cref{steps} shows the performance on the ICLT task according to the number of training steps. The result is consistent with the trend in other domains that scaling up the training dataset leads to a strict performance improvement on the downstream tasks~\cite{NEURIPS2020_1457c0d6, kaplan2020scaling, zhai2021scaling}. Interestingly, we can observe the considerable positive effects of batch shuffling. It is also surprising that CLUE trained on $10\%$ of the dataset---using only 1,130,000 users---with random shuffling can achieve competitive results with the LightGCN trained on the full dataset of historical logs on the ICLT task. From the results of 
\cref{batch-dataset-model-comp},~\ref{steps}, and~\ref{transfer}, we can conclude that all four factors must scale up in tandem for optimal performance. 
\setlength{\tabcolsep}{13.3pt}
\ctable[
    caption = {Performance comparison of Single and Stacking Transformer encoder for CLUE. They both compete with the same GPU resources.
},
    label = tab:architecture,
      doinside=\small
]{cccc}{
}{

\toprule Tasks & Metrics & Single & Stacking \\
\midrule
     \multirow{3}{*}{PCR} & HR@10 & 0.5341 & \underline{0.5414}  \\
     & NDCG@10 & 0.7377 & \underline{0.7418}  \\
     & MRR & 0.6808 & \underline{0.6857} \\
     \midrule
     \multirow{3}{*}{ICLT} & HR@10 & 0.4956 & \underline{0.5081}  \\
     & NDCG@10 & 0.6934 & \underline{0.7036}   \\
     & MRR & 0.6333  & \underline{0.6440}    \\
\bottomrule
}

\setlength{\tabcolsep}{8.0pt}
\ctable[
    caption = {
    Decreasing the output feature dimension of CLUE does not lead to a significant difference in the performance on the downstream (PCR) tasks.
    },
    label = tab:dimension,
    pos=t,
      doinside=\small
]{cccc}{
}{
\toprule
Output Dimensions & HR@10 & NDCG@10 & MRR  \\
\midrule
300D & 0.5414 & 0.7418 & 0.6857  \\
2160D & 0.5360 & 0.7390 & 0.6822  \\
\bottomrule
}

\noindent\textbf{Generalization on Other Data Distributions.}
The in-distribution test loss as a function of computation is shown in \cref{transfer}-Left. We empirically verify that the loss on the pretraining dataset improves smoothly with more computation. 
This result aligns with the reports in other domains, which has shown that increasing the amount of computation positively affects the performance of pretrained models~\cite{NEURIPS2020_1457c0d6, kaplan2020scaling, zhai2021scaling}. Moreover, when we transfer knowledge to datasets with a different distribution than the one used during pretraining, the resulting test losses show a strong correlation with the pretraining performance. These results indicate that generalization ability to various data distributions is strongly dependent on the pretraining test loss (\cref{transfer}-Right).

\subsection{Ablation Study on Encoder Architectures}
\noindent\textbf{Efficacy of Stacked Transformers.}
We conduct an ablation study to analyze the effects of stacking Transformers, i.e, the Item Transformer and the Service Transformer, in our model. The stacked model (original CLUE) first extracts the item embedding from the Item Transformer then passes them to the Service Transformer to compute the user embeddings. On the other hand, the single model computes user embeddings from a single Transformer; has no module to extract item embeddings. For fair comparison, we train a single Transformer encoder with 235M parameters, sequence length 2,048, batch size 256, and the same GPU resources as the best CLUE model. As presented in \cref{tab:architecture}, the proposed stacking approach performs better than the single Transformer encoder with the same configuration. We conjecture that separating the encoding process enhances the representation quality of the encoder. Furthermore, the stacking approach allows our model to observe more user behaviors compared to using a single Transformer.

\noindent\textbf{Output Dimensions.}
The final user feature dimensions of the non-curated best CLUE model is 2,160. The 2,160 feature dimensions for whole users results in a size of 153 GB when stored in half-precision floating-point format. If we slightly increase the loading time or the number of services, the storage becomes too large to handle. Thus, we conduct an experiment on whether the output dimension of CLUE affects its performance. We added a single MLP layer for reducing the dimension of the encoder outputs. \cref{tab:dimension} shows that decreasing the output dimension does not lead to any performance degradation on the downstream transfer tasks.

\section*{Limitations}
\begin{figure}[t]
\begin{center}
\centerline{\includegraphics[width=0.80\columnwidth]{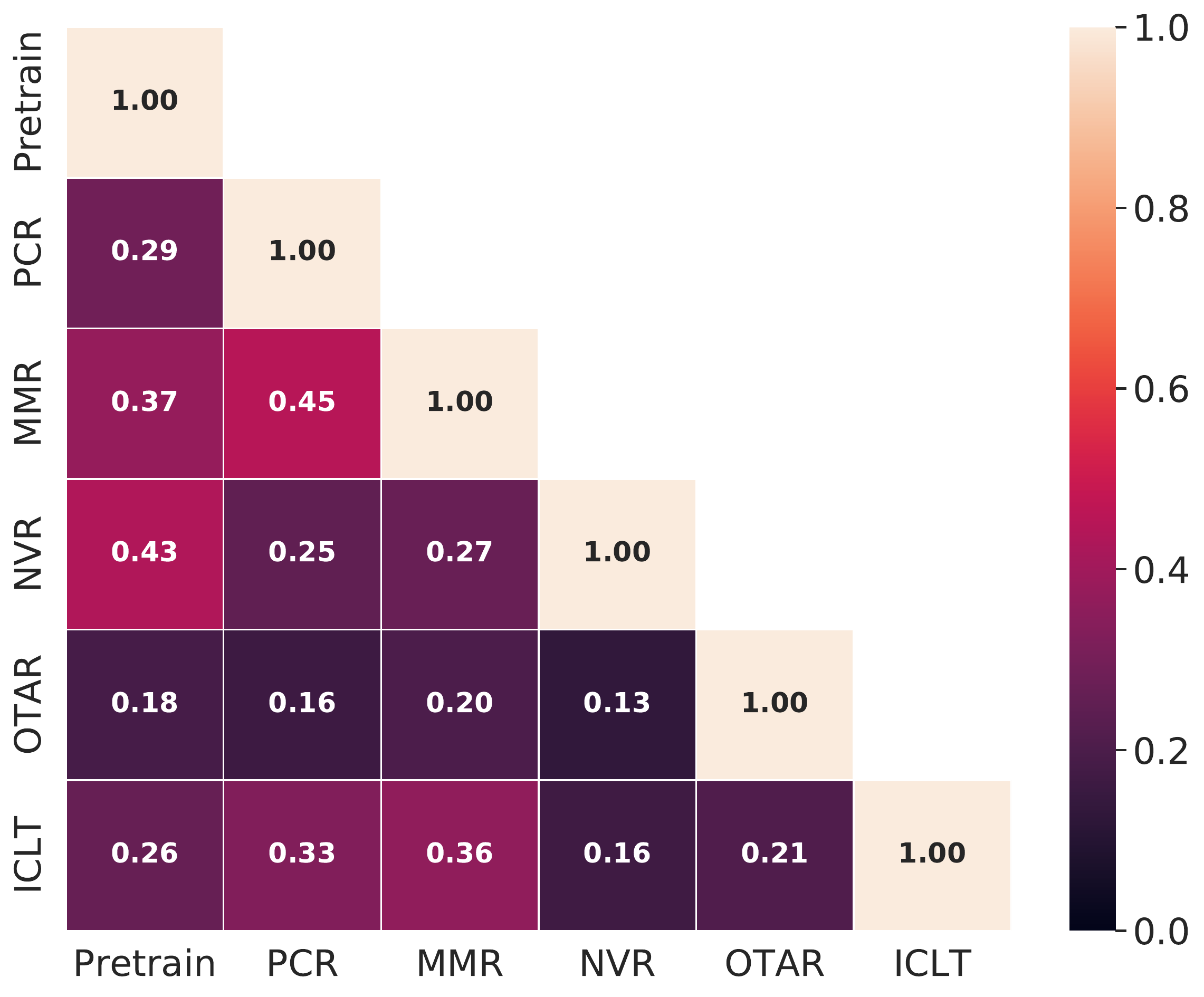}}
\caption{The Kendall rank correlation of the token distribution between the pretraining and downstream domains. We use the user behavior logs $u$ to calculate the token distributions.} 
\label{correlation}
\end{center}
\end{figure} 
Although we have adopted a scaling-up strategy, we believe that a one-size-fits-all approach to the user modeling area continues to be challenging, and there remains tricky downstream tasks that have not yet been addressed.
The domain of the pretrain data can be a constraint in learning a complete general-purpose user representation.
Thus, we investigate the  similarity of source and target domain with performance on the target task.
Recently, \citet{gururangan2020don} investigated whether it is helpful to transfer a pretrained model to the domain of a target task. They used the vocabulary overlap ($\%$) to measure the similarities between the source and the target domains. Since it is difficult to measure the vocabulary overlap of the user behavior dataset between different domains, we estimate the Kendall rank correlation~\cite{abdi2007kendall} using the token distribution of the user behavior logs of each domain.

\cref{correlation} shows the correlation across the downstream task datasets. We observe that the correlation and performance improvement do not align perfectly. According to the results of \cref{tab:downstream-tasks} and \cref{correlation}, the correlation and the relative performance increase of CLUE against the task-specific models---except the pretraining models---show a trend, but not in all cases. Our work has not made much progress towards finding a criteria for a well-transferable domain. A more careful study on this subject is left for future research.

\section*{Related Work}
\textbf{Contrastive Learning.}
Contrastive learning~\cite{hadsell2006dimensionality} aims to learn high-quality representations by contrasting positive sample pairs against negative sample pairs.~\citet{chen2020simple} and \citet{caron2020unsupervised} demonstrated that contrastive visual representation learning could produce  comparable results to the supervised method on several vision tasks. These approaches have been popularized for other domains, such as language modeling~\cite{gao2021simcse} and speech recognition~\cite{baevski2020wav2vec}. Recently, \citet{akbari2021vatt} and \citet{radford2021learning} presented a method for learning a flexible prediction space for different modalities by employing contrastive losses to train the model. 

\noindent\textbf{Scaling Law.} 
Over the decades, the connection between scaling and generalization has been studied broadly from both theoretical and empirical perspectives~\cite{mhaskar1996neural, NEURIPS2020_1457c0d6, kaplan2020scaling, bahri2021explaining, zhai2021scaling, hutter2021learning}.
Several studies theoretically demonstrated that the training error scales as a power-law with larger model capacity and more data~\cite{mhaskar1996neural, bahri2021explaining, hutter2021learning}. Recent research focuses on empirical analysis of scaling laws for real-world applications~\cite{NEURIPS2020_1457c0d6, kaplan2020scaling, zhai2021scaling}. These works showed that scaling up the model and dataset size is a promising approach for achieving outstanding performances in several NLP and computer vision tasks, e.g., machine translation, generative modeling and image recognition.

\noindent\textbf{General-Purpose User Representation Learning.} Compared to task-specific user representation learning, the research about the general-purpose user representation learning is still in its early stage.
\citet{yuan2021one} continually learn user representations on multiple tasks with a single model without network expansion and catastrophic forgetting. 
\citet{wu2021userbert} and \citet{hou2022towards} contrastively pretrain user models to effectively capture the relations between user behaviors and inherent user interests.
Note that these models are all restricted to fine-tuning based approaches~\cite{yuan2021one, wu2021userbert, hou2022towards}. 
The recent progress in general representation learning naturally attempts to build more efficient and widely adaptable user representations.
\citet{gu2021exploiting} employ a novel objective function named behavioral consistency loss to preserve the user’s long-term interest and benefit from diverse behaviors. 
\citet{sun2021interest} present interest-oriented contrastive learning, which maximizes the agreement between short- and long-term interest representations of the same users.
These methods can be applicable to numerous downstream tasks without further modifications~\cite{gu2021exploiting, sun2021interest}.

Following the previous studies, we focus on learning adaptable, general user representations and connecting the bridge between user representation learning and the empirical scaling law.

\section*{Conclusion and Impact}
We explore whether the success of task-agnostic pretraining in other domains is valid in user modeling areas.
We present CLUE trained on the billion scale real-world user behavior data to learn general-purpose user representations. We benchmark the various downstream tasks with a simple MLP, and achieve promising results, including company-level transferring task.
We further investigate the empirical scaling laws and the generalization ability of our method, and find that the power-law learning curve as a function of computation (PF-days) is observed in the experiments.
Despite the remaining limitations, we believe that our empirical analysis can share useful insights on large-scale learning for user representations with user modeling and recommender system communities.

\section*{Acknowledgments}
The authors would like to thank the NAVER CLOVA ML X team for insightful comments and discussions. We would also like to thank the NAVER Smart Machine Learning (NSML) platform team~\cite{sung2017nsml, kim2018nsml} for their critical work on the software and hardware infrastructure on which all the experiments were performed.
\bibliography{main}
\appendix
\section{Datasets} \label{sec:app-dataset}
\setlength{\tabcolsep}{6pt}
\ctable[
    caption = {Statistics of the Training Dataset.},
    label = tab:data,
    pos=ht,
 	doinside=\normalsize
]{cc}{
}{
\toprule
\textbf{Contents} & \textbf{Values} \\
\midrule
\textbf{Periods} & Oct, 2018 - Oct, 2020 \\
\textbf{Number of users}      & 11,294,122  \\
\textbf{Number of behaviors}     & 5,313,992,425  \\
\textbf{Number of behavior tokens} & 50,741,643,225        \\
\textbf{Maximum token length}      & 32  \\
\bottomrule
}
\subsection{Synthetic User Behavior Dataset Examples}
\begin{tcolorbox}[
title={Item description},
height=7.6cm,
halign=left,
valign=top
]
\textbf{Search query} \\
Mapo Han River Metro Xi $\rightarrow$ Pororo Camping Car $\rightarrow$ Geomdan Kumho New Town $\rightarrow$ Squid Game $\rightarrow$ Secret Forest Broadcast Time $\rightarrow$ Bookshelf $\rightarrow$ Home Surveillance Camera $\rightarrow$ Mapo Ehak Galbi 
\begin{align*}
\end{align*}
\textbf{Purchased product description}\\
Latex Powder Free Gloves $\rightarrow$ Conte quilted Vest zip up $\rightarrow$ Multi-use Super Silicone Broom $\rightarrow$ Children Kids Junior Basic T-Shirts $\rightarrow$ Comfortable Corduroy Leggings Pants for Weekly $\rightarrow$ High Quality Expert Olive Oil $\rightarrow$ Illy Coffee Capsules $\rightarrow$ Children's Bedroom Slippers Chandelier 
\end{tcolorbox}
\begin{tcolorbox}[
title={Item tokens},
height=10.8cm,
halign=left,
valign=top
]
\textbf{Search query tokens}
\begin{align*}
&\text{[[26521, 22276, 38343, 12161,\,\:\:\:\:\:\:\:\:\:\,0],} \\
&\text{\,\,[\:\:3097, 15940, 10626,\:\:\:\:\:\:653,     \:\:\:\:\:\:\:\:\:0],}  \\
&\text{\,\,[\:\:1658,   \:\:\:\:\,598,  \:\:8999, 32855,     \:\:\:\:\:\:\:\:\:0],}  \\
&\text{\,\,[16499,  \:\:\:\:\,303,  \:\:4334,  \:\:5118,  \;\;1371],}  \\
&\text{\,\,[35858,     \:\:\:\:\:\:\:\:\:0,     \:\:\:\:\:\:\:\:\:0,     \:\:\:\:\:\:\:\:\:0,     \:\:\:\:\:\:\:\:\:0],}  \\
&\text{\,\,[\:\:1568,   \:\:\:\:\,891,   \:\:\:\:\,742,     \:\:\:\:\:\:\:\:\:0,     \:\:\:\:\:\:\:\:\:0],}  \\
&\text{\,\,[\:\:6759,   \:\:\:\:\,347,   \:\:\:\:\:\:\:67, 30617,     \:\:\:\:\:\:\:\:\:0],}  \\
&\text{\,\,[26521,   \:\:\:\:\,264,   \:\:\:\:\,643,  \:\:4109,     \:\:\:\:\:\:\:\:\:0]]}\\
\end{align*}
\textbf{Purchased product description tokens}
\begin{align*}
&\text{[[\:\:\:\:\,333,  \:\:9874, 20235, 14771,     \:\:\:\:\:\:\:\:\:0,     \:\:\:\:\:\:\:\:\:0],} \\
&\text{\,\,[16436, 25190,   \:\:\:\:356,   \:\:\:\:\,676,  \;\;7121, 20382],}  \\
&\text{\,\,[12666,  \:\;3047,  \:\:1567,  \:\:5076, \,16910,     \:\:\:\:\:\:\:\:\:0],}  \\
&\text{\,\,[12859, 11709, 19096,  \:\:5936, \:\:\:\:\,810,\:\:\:\:\:\:\:\:\:\:\,0],}  \\
&\text{\,\,[\:\:\:\:\,590,  \:\:8746,  \:\;1593,   \:\:\:\:\,291, 29324, 17274],}  \\
&\text{\,\,[\:\:\,9545, \:\:8866, \:\:2966,   \:\:\:\:\,290,   \:\:\;\;422, \:\:4537],}  \\
&\text{\,\,[\:\:\,4909, 26935,  \:\:4479,\:\:\:\:\:\:\:\:\:\:\,0,\:\:\:\:\:\:\:\:\;\;0,\:\:\:\:\:\:\:\:\:\;0],}  \\
&\text{\,\,[23035, \,24403,   \:\:\:\:350, \:\!12859, 11975, 13475]]}  \\
\end{align*}
\end{tcolorbox}

\section{Training Details} \label{sec:app-training}
\setlength{\tabcolsep}{15pt}
\ctable[
    caption = {Best CLUE model hyperparameters. The hyperparameters are same for both Item and Service Transformer. The micro-batch size means the batch size of 1 GPU. },
    label = tab:hyperparams,
    pos=ht,
 	doinside=\normalsize
]{cc}{
}{
\toprule
\textbf{Hyperparameter} & \textbf{Value} \\
\midrule
\textbf{Vocabulary size}      & $50,257$  \\
\textbf{Micro-batch size}     & $4$  \\
\textbf{Global-batch size}      & $256$  \\
\textbf{Max sequence length}     & $512$  \\
\textbf{Embedding dimension} & $720$        \\
\textbf{Feedforward network dimension} & $2,880$ \\
\textbf{Layers} & $8$ \\
\textbf{Heads} & $6$  \\
\textbf{Dropout rate} & $0.1$ \\
\textbf{Adam $\beta_{1}$} & $0.9$ \\
\textbf{Adam $\beta_{2}$} & $0.98$ \\
\textbf{Adam $\epsilon$} & $10^{-6}$ \\
\bottomrule
}
We use the Zero Redundancy Optimizer~\cite{rajbhandari2020zero} with AdamW~\cite{loshchilov2017decoupled} and weight decay regularization applied to all weights with a decay rate of $0.1$. We update the model using an initial learning rate of $0.0005$, then incorporate learning rate warm-up over the first $1\%$ steps, followed by cosine decay~\cite{loshchilov2016sgdr} to lower the learning rate to $10\%$ of its value. The learnable temperature parameter $\tau$ is initialized to $14.27$ and its value is clipped to prevent scaling the logits by more than $100$. We use the automatic mixed-precision~\cite{micikevicius2017mixed} package in Pytorch~\cite{paszke2019pytorch} to accelerate training and reduce memory usage. Gradient norm clipping~\cite{pascanu2013difficulty} is used with the max norm set to $0.01$ to stabilize training. The calculation of the embedding similarities is distributed across a multi-node cluster. Then, all the shared similarities are used for computing the logits, but only the subset of the pairwise similarities residing on an individual GPU is used for the gradient updates on that GPU. 
We shuffle the dataset at every epoch and train the model for 8 epochs, where the transfer performance begins to plateau. 
The total training time takes 7 days on $64$ V100 GPUs. Unless otherwise specified, all results reported in this paper as “CLUE” use this model, which we find to perform the best. The hyperparameters for the best CLUE model is detailed in \cref{tab:hyperparams}.

\section{Inter-company-level transfer framework}  \label{sec:app-iclt}
\begin{figure}[ht]
\includegraphics[width=1.0\columnwidth]{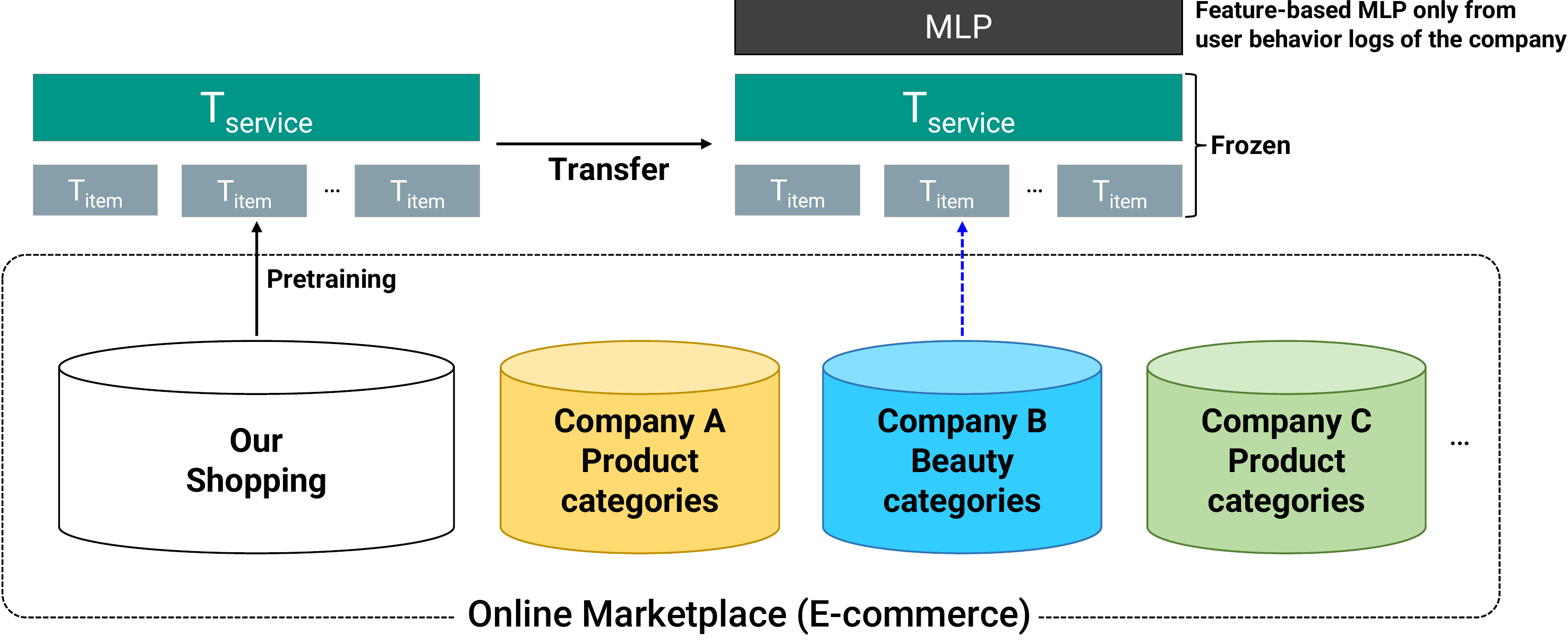}
\caption{Overall framework of the Inter-Company-Level-Transfer task. CLUE pretrained on our shopping data provides downstream user representations extracted from the user behavior logs of other companies. Such user representations are then projected by a simple MLP for the downstream task.}
\label{fig_itlc}
\end{figure} 
We investigate the company-level transferability of CLUE by validating on the ICLT task. To achieve this, we collect user logs from a beauty service provided by another company on the same open market platform as ours but operated independently as shown in \cref{fig_itlc}. We apply a feature-based approach with a simple MLP using user features from the target company data.

\section{Downstream Tasks} \label{sec:app-downstream}
\subsection{Benchmark Tasks}
We collect 243,100 users who have review logs on both \textit{“Books”} and \textit{“Clothing Shoes and Jewelry”} categories of Amazon review dataset~\cite{ni2019justifying}. The 143,100 users are used for pretraining datasets, and the remaining users are used for downstream tasks. The product titles are used as the items.

\noindent\textbf{Books}: We collect 1,298,489 review logs of 100,000 unique users and 504,572 unique books.

\noindent\textbf{Clothing}: We collect 928,598 review logs of 100,000 unique users and 314,943 unique clothing, shoes, and jewelry.
\subsection{Industrial Tasks}
\noindent\textbf{Product Collection Recommendation (PCR)}: Product Collection is a collection of products designed by merchandisers, and has a collection-specific description such as  ``Best DVDs for babies'', ``Big sale on graphic cards'', and ``The most profitable products''. Each collection is represented by a banner that is linked to a page displaying a list of products. The task is to recommend this banner properly. We collect 590,770 click logs of these banners containing 300,000 unique users and 8,652 unique Product Collections. The banner titles are used as the the items.

\noindent\textbf{Marketing Message Recommendation (MMR)}: Chatbot marketing is a strategy for promoting user engagement to generate sales. The chatbot sends personalized events and products to users via messages. We collect 502,362 click logs of 300,000 unique users and 17,530 unique messages. We use these messages as the items.

\noindent\textbf{News View Recommendation (NVR)}: An online newspaper is served either in the form of a stand-alone publication or as the online version of a printed periodical. We collect 4,029,661 news views of 299,588 unique users and 293,834 news articles. The news articles cover various topics including, but not limited to: politics, health, technology, business, entertainment and sports. We use the title of the news articles as the items.

\noindent\textbf{Online Travel Agency Recommendation (OTAR)}: The online travel agency is a leisure platform that provides online travel accommodations and travel services around the world.
We collect 177,281 reservations of 142,051 unique users from 2,485 accommodations. The task is to recommend the hotels to the travellers. We use the hotel name as the item.

\noindent\textbf{Favorite Webtoon Recommendation (FWR)}: Webtoons are animated cartoons or comics that are published online. We collect 4,323,578 favorited webtoons from 296,469 unique users and 1,573 unique webtoons. We use the synopsis of the webtoons as item.\footnote{We exclude task-specific supervision learning for FWR because this dataset is collected without date information.} 

\noindent\textbf{Inter-Company-Level Transfer (ICLT)}: Existing user models are mainly pretrained from service-level or intra-company data. If a model is inter-company-level transferable, the model can provide much more generalized services across multiple platforms and companies. The task is to validate if a model pretrained on specific company data by CLUE can generate effective user representations from user behavior logs of another company. To the best of our knowledge, this is the first study on the possibility of inter-company-level transferability in recommender systems. We collect 179,435 users and 13,841 items from the service of the target company, beauty and cosmetic e-commerce platform. The total amount of purchase logs is 558,992. The title of the product is used as the item. 

\section{Comparison Models} \label{sec:app-comparison}
To verify the effectiveness of CLUE, we compare it with the following downstream models:

\noindent\textbf{DeepFM (Task-specific):} DeepFM~\cite{guo2017deepfm} is one of the most popular methods in recent recommender systems. It integrates the architectures of FM~\cite{rendle2010factorization} and deep neural networks by using both low- and high-order feature interactions.

\noindent\textbf{BST (Task-specific):} Behavior Sequence Transformer~\cite{chen2019behavior} embeds task-specific historical logs as low-dimensional vectors, which are passed to the transformer layers to model the sequential signals underlying the users’ behaviors. 

\noindent\textbf{LightGCN (Task-specific):} 
LightGCN~\cite{he2020lightgcn} discards the unnecessary components of a Graph Convolution Network~\citep{kipf2016semi} for collaborative filtering. It trains user and item embeddings by linearly propagating them on a bipartite interaction graph. The weighted sum of the embeddings propagated at each layer is used as the final embedding.

\noindent\textbf{YTMoE (Task-specific):} 
YTMoE~\cite{zhao2019recommending} combine several tasks from Click-Through-Rate (CTR) prediction and demonstrates the generality of learned representation on a large-scale multi-objective ranking system.

\noindent\textbf{UserBERT (Task-specific):} 
UserBERT~\cite{wu2021userbert} incorporates two self-supervision tasks, Masked Behavior Prediction (MBP) and Behavior Sequence Matching (BSM). These two contrastive pretext tasks effectively capture the relations between user behaviors and inherent user interests. It finally finetuned models on target tasks.

\noindent\textbf{UniSRec (Task-specific):} 
UniSRec~\cite{hou2022towards} propose cross-domain and cross-platform user representation models. It combines parametric whitening and MoE adaptor architectures to learn user representation. UniSRec pretrains models by sequence-sequence contrastive learning of same users and then finetuned models to target tasks.

\noindent\textbf{ShopperBERT (Task-agnostic):} ShopperBERT~\cite{shin2021one4all} is a pretrained model that leverages BERT's framework for learning general-purpose user representations. It trains embeddings through a pretext task that uses product categories as tokens and predicts the masked information under the [MASK] token.

\noindent\textbf{SimCLR (Task-agnostic):} Inspired by the success of SimCLR~\cite{chen2020simple} in computer vision, this setup follows the same architecture as CLUE, but has a different contrastive objective. We create positive pairs by augmenting a given user behavior sequence in two different ways, selected among cropping, masking, and re-ordering the tokens.

\noindent\textbf{Hybrid:} Hybrid combines the user feature by CLUE with the final output vectors of \textbf{task-specific} models. We expect the pretrained user representations to greatly enhance the performance of task-specific models in many downstream tasks. We choose the best-performing model between BST and LightGCN for each downstream task.  

\end{document}